\documentclass[12pt]{article}

\input epsf
\usepackage{amssymb,wrapfig}
\usepackage[matrix,arrow,curve]{xy}
\usepackage{cite}
\usepackage{graphicx}

\makeatletter
\@addtoreset{equation}{section}
\makeatother

\def\nn {{\cal N}}

\def\cc {{\Bbb C}}
\def\pp {{\Bbb P}}
\def\zz {{\Bbb Z}}
\def\del {\partial}
\def\cy {Calabi--Yau}
\def\ka {K\"ahler}
\def\del {\partial}

\def\stt {{$\mathrm{SU(3)}\times\mathrm{SU(3)}$}}
\def\sut {{${\rm SU}(3)$}}
\def\vol {\mathrm{vol}}

\makeatletter
\@addtoreset{equation}{section}
\makeatother

\newcommand{\nc}{\newcommand}
\nc{\eps}{\epsilon}

\begin{document}

	\begin{titlepage}
	\begin{center}

	\vskip .3in \noindent


	{\Large \bf{New string vacua from twistor spaces}}

	\bigskip

	 Alessandro Tomasiello\\

	\bigskip
	Jefferson Physical Laboratory, Harvard University, 
	Cambridge, MA 02138, USA 

	\vskip .5in
	{\bf Abstract }
	\vskip .1in
	\end{center}

	\noindent 
We find a new family of AdS$_4$ vacua in IIA string theory. The internal
space is topologically either the complex projective space $\cc\pp^3$ 
or the ``flag manifold'' ${\rm SU}(3)/({\rm U}(1)\times {\rm U}(1))$, but the metric is in
general neither Einstein nor \ka. All known moduli are stabilized
by fluxes, without using quantum effects or orientifold planes. The
analysis is completely ten--dimensional and does not rely on
assumptions about Kaluza--Klein reduction.     	
	\vfill
	\eject


	\end{titlepage}

\section{Introduction} 
\label{sec:intro}

The search for realistic string vacua usually
proceeds in steps, by mixing and matching different features
of the theory. For example, achieving a positive cosmological constant $\Lambda$ is not easy 
\cite{maldacena-nunez}. For that reason, one usually starts with
a vacuum with a vanishing or negative $\Lambda$, and then attempts
to modify it using ingredients that evade the no--go argument in 
\cite{maldacena-nunez}. 

One popular construction of this kind \cite{kklt} actually proceeds
in three steps. It starts from a class of Minkowski vacua 
\cite{dasgupta-rajesh-sethi,grana-polchinski,
giddings-kachru-polchinski};
it then gives mass to the only massless scalar in those vacua, by using 
quantum corrections, and in the process introducing a {\it negative}
$\Lambda$; it finally makes $\Lambda$ positive, and breaks 
supersymmetry, by using brane--antibrane pairs. Even if this
construction seems to produce very large numbers of vacua with
pleasing features, it should encourage us to look further and to 
ask whether there are different families of vacua, or maybe different
constructions, that can later be made realistic. For example, the Minkowski vacua from which \cite{kklt} starts are obtained by
compactifying IIB on \cy\ manifolds. One can ask if there are
other manifolds that give supersymmetric flux vacua; the answer
turns out to be remarkably simple \cite{gmpt2,t-reform}, and 
to involve interesting geometrical concepts. A case study, however,
shows \cite{gmpt3} that finding concrete examples to the
conditions in \cite{gmpt2} is slower than one would like. 

Other than this geometrical simplicity, there is no reason, however, 
not to start with $\Lambda$ negative to begin with. There are indeed
several examples of supersymmetric AdS$_4$ with moduli stabilized. 
For example, \cite{dewolfe-giryavets-kachru-taylor} find a simple
IIA family with no moduli and in which possible corrections are parametrically under control, using orientifold planes and again \cy\ manifolds. As later shown in \cite{acharya-benini-valandro}, 
from a purely ten--dimensional
perspective these vacua are found by using a low--energy approximation
in which the orientifold sources are effectively smeared. Although I
think this approximation is correct, there is no reason one should
use orientifold planes at all to find AdS$_4$ vacua (whereas they
are a prominent way of evading the arguments in \cite{maldacena-nunez}
for Minkowski vacua). It should then be possible to find many
vacua even without them.

The oldest construction of AdS$_4$ vacua in IIA
 is from M--theory via
the so--called Freund--Rubin choice of fluxes (for a review see \cite{duff-nilsson-pope}). The internal space $M_7$ is, in this case, 
an Einstein manifold. Some of those vacua can also be reduced
to (or directly found in) IIA. For example, AdS$_4\times S^7$
in M--theory can also be understood dually as AdS$_4\times \cc\pp^3$
in IIA with $F_2$ flux \cite{nilsson-pope,watamura-2,sorokin-tkach-volkov-11-10} (although some subtlety about the amount
of perturbative supersymmetry arise because the reduction does
not preserve all the supercharges \cite{duff-lu-pope}). Another 
embarrassingly simple IIA construction was found in \cite{behrndt-cvetic,behrndt-cvetic2}. The idea is to
consider metrics which are not \cy, but whose deviations from the
\cy\ condition is (in a sense to be reviewed later, in terms of
an internal ``\sut\  structure'') parameterized by a single
real number $W_1$. These metrics are called {\it nearly \ka }.
They are also Einstein,
the scalar curvature being  proportional to $|W_1|^2$. 
By a suitable choice of the internal fluxes (i.~e.~by taking them 
to be singlets under the internal \sut\ structure), all the supersymmetry equations then reduce to easily solvable algebraic equations involving  scalars.\footnote{Nearly \ka\ manifolds have
also appeared in heterotic string theory \cite{manousselis-prezas-zoupanos,frey-lippert}.} 
It turns out that a
nearly \ka\ metric exists on $\cc\pp^3$; it is different from the
usual Fubini--Study metric.

In this paper I will generalize both of these two constructions
of vacua on AdS$_4\times \cc\pp^3$, in a way that in a sense
interpolates between the constructions in \cite{nilsson-pope,watamura-2,sorokin-tkach-volkov-11-10}
and \cite{behrndt-cvetic,behrndt-cvetic2}. These metrics are in general not
Einstein and in particular not nearly \ka, nor \ka. (Not surprisingly
to flux compactifications aficionados, the almost complex structure
is not integrable, because of the cosmological constant.) Nor are
the fluxes simply singlets of the internal \sut\ structure. 

The way I found these vacua is by considering
$\cc\pp^3$ as a twistor fibration (that has fiber $S^2$) on $S^4$, with a slightly unusual choice of non--integrable
almost complex structure that turns out to have vanishing $c_1$. 
The metrics are obtained by varying the relative factor between
the metric on the fiber and the one on the base; one can think of it
as of a ``squashing parameter''. 
The construction can be repeated with few changes
for the twistor space ${\rm SU}(3)/({\rm U}(1)\times {\rm U}(1))$ of $\cc\pp^2$, but we will focus mostly on $\cc\pp^3$. 

There are infinitely many of these vacua; because of flux quantization, 
all the known moduli are stabilized. In a sense, instead of starting
with many geometrical moduli and finding then a way of stabilizing
them (like one does with \cy\ manifolds), we start with a space that has very few moduli to begin with. 
Just like in \cite{dewolfe-giryavets-kachru-taylor}, dilaton and
internal curvature can be made parametrically small.

Given that our computations are always purely ten--dimensional, we have nothing to say in this paper about the
low--energy effective action describing excitations around these 
vacua.\footnote{For the vacua in \cite{behrndt-cvetic,behrndt-cvetic2}, some of which are a special 
case of those presented here, an effective theory with $\nn=2$
supersymmetry was recently proposed in \cite{kashanipoor}.
For a similar $\nn=1$ analysis on ${\rm SU}(3)/{\rm U}(1)\times {\rm U}(1)$ see \cite{house-palti}; for another nearly \ka\ space, 
$S^3\times S^3/\zz_2^3$, see \cite{aldazabal-font}.}
It should not be difficult, however, if need be, to compute these
effective theories, perhaps using an alternative description of
these metrics in terms of group cosets \cite{ziller}. 

These examples also illustrate a limitation inherent to the usual approach of finding 
first an effective theory by KK reducing on a space, and finding then vacua for this effective theory. While this looks physically
very reasonable, KK reducing on a general manifold
is in fact not easy, in general: so far, the only examples fully
understood are \cy's, parallelizable manifolds (like the so--called ``twisted tori'', used in Scherk--Schwarz constructions\footnote{These have been used to argue for $AdS_4$ vacua for example in \cite{villadoro-zwirner,camara-font-ibanez,
derendinger-kounnas-petropoulos-zwirner}}) or cosets. Proposals exist on how to understand
more general manifolds (see for example \cite{grana-louis-waldram}), 
but they are plagued by many geometrical issues \cite{kashanipoor-minasian}, 
which so far seem to be under control only in simple cases \cite{gurrieri-louis-micu-waldram,kashanipoor} (although one
can show that four-- and ten--dimensional supersymmetry are
equivalent \cite{koerber-martucci-ten-four,cassani-bilal}). 
Given this state of affairs, one might want to look for vacua first, 
and only later for effective theories. 

On a different note, it would be interesting to know the CFT
duals to these vacua; as remarked in \cite{schwarz-chernsimons}, the Romans mass
$F_0$ should give rise to a Chern--Simons theory, perhaps of the
type discussed in \cite{gaiotto-yin}.

After reviewing in section \ref{sec:review} the conditions (\ref{eq:susy}) and (\ref{eq:dw2})
for supersymmetric vacua, and the geometrical information we
need in section \ref{sec:geo}, 
we will show the existence of the new vacua in section \ref{sec:finding}.


\section{Review of Anti--de Sitter vacua in type IIA} 
\label{sec:review}

We will start by reviewing the conditions imposed by supersymmetry
on the internal geometry, and by specializing them to the case in
which no sources (branes or orientifold planes) are present.

This computation has been carried out in \cite{lust-tsimpis}; in \cite[Sec.~7]{gmpt3} it has been rederived using the techniques of 
\stt\ structures. Since the last presentation seems smoother to me
(no doubt because of personal bias), I will use the notation 
in \cite{gmpt3} (save for one minor difference to be noted later). Rather than reviewing here the machinery of 
generalized complex structures, I will cut to the chase and
describe the final result of that analysis in terms of (hopefully)
lighter mathematics.

We do need, however, the concept of an \sut\ structure. This is
just the type of structure that we are familiar with from \cy\ manifolds, but without the differential equations. Namely, 
an \sut\ structure is a pair of forms $(J, \Omega)$ such that 
\begin{itemize}

\item $J$ is a real two--form, $\Omega$ is a complex three--form and {\it decomposable} (locally the wedge product of three complex
one--forms); $\Omega$ then determines an almost complex
structure $I$;

\item $\frac 43 J^3=i \Omega\wedge \bar \Omega\neq 0 $ everywhere;

\item $J\wedge \Omega=0$;

\item the tensor $g=JI$ (which is symmetric because of the 
conditions above) is positive definite. 
\end{itemize}

Notice that with respect to $I$, it is easy
to see from the above conditions that $J$ is $(1,1)$ and 
$\Omega$ is $(3,0)$.

A \cy\ manifold can be defined then as a manifold on which 
\begin{equation}\label{eq:cy}
	dJ=0=d \Omega \qquad {\rm (Calabi-Yau).}
\end{equation}
In this paper we are not interested in \cy\ manifolds, however; 
and we will
shortly see why. 

The supersymmetry 
conditions in IIA for an AdS$_4$ vacuum {\it with \sut\ structure}
read\footnote{One of
the results of \cite{gmpt3} is that the warping factor has to be
constant. One can then eliminate it completely from the equations
by using $\phi=3A$ (see (7.9) in \cite{gmpt3}) and by redefining
the parameters as $m_{\rm here}=m_{\rm there}e^{-A}$, 
$\tilde m_{\rm here}=\tilde m_{\rm there}e^{-A}$ }. 
\begin{equation}\label{eq:susy}
	\begin{array}{c}\vspace{.4cm}
		dJ = 2\tilde m {\rm Re} \Omega\ ,\qquad
		d \Omega= i(W_2^- \wedge J -\frac43 \tilde m J^2)
		\ ,\qquad H= 2m {\rm Re} \Omega
		\ ;\\
		g_s F_0=5m \ ,\qquad g_s F_2=-W_2^-+\frac13 \tilde m J
		\ ,\qquad g_s F_4 = \frac32 m J^2
		\ ,\qquad g_s F_6=-\frac12\tilde m J^3\ .
			\end{array}
\end{equation}
Here, $m$ and $\tilde m$ are two real numbers; $W_2^-$
is a {\it primitive} $(1,1)$--form 
(the strange notation comes from \cite{gray-hervella,chiossi-salamon}; primitive means that $W_2^-\wedge J^2=0$); $g_s$ is the constant string coupling. Notice that the parameter $m$ is related, but not exactly 
equal, to the Romans mass $F_0$. The cosmological constant 
in four dimensions is given by 
\begin{equation}\label{eq:cc}
	\Lambda= - 3 (m^2 + \tilde m^2) \ . 
\end{equation}

In (\ref{eq:susy}) the $F_i$'s are the {\it internal } fluxes.
There are also ``external'' fluxes, that span the AdS$_4$
directions as well as some of the internal directions; these
are determined by the internal fluxes by ten--dimensional Hodge
duality. For example, there is also a flux extended along the 
AdS$_4$ directions only: 
\begin{equation}
	F^{\rm ext}_4= *_{10} F_6= \vol_4 *_6 F_6 = \frac{3\tilde m}{g_s}\vol_4\ .
\end{equation}
For the same reason, there are also fluxes of the form $\vol_4$
wedge an internal two--form, four--form and six--form.
We will never mention again the external fluxes $F^{\rm ext}_i$;
we will always use the internal $F^{\rm int}_i\equiv F_i$. 

If one wants to attach a name to the geometrical part of (\ref{eq:susy}), one could say that they describe a ``half--flat''
manifold (as also noticed in \cite{acharya-benini-valandro}), 
namely one such that $d {\rm Re} \Omega=0=d J^2$; 
although it is a very particular one, and hence the name is probably
not very useful. Notice also that, even for more general 
solutions with \stt\ structure, one still gets that all vacua
are ``generalized half--flat'' manifolds, as explained in \cite{gmpt3}.
(We will see at the end of this paper how that more general analysis
should be relevant for the vacua in \cite{dewolfe-giryavets-kachru-taylor}.)

In any case, the supersymmetry conditions have to be supplemented by
the Bianchi identities for the fluxes. If we impose that there 
are no sources, these read 
\begin{equation}
	\label{eq:bianchi} d F_k = H \wedge F_{k-2}\ . 
\end{equation}
In fact, we have already used the $k=0$ case, $d F_0=0$: 
this is how it was derived in \cite{gmpt3} that 
the warping $A$ and the dilaton $\phi$ must be constant.  
One would also need, a priori, to impose the equations of motion
for the fluxes, $d *_6F_k=- H \wedge *_6 F_{k+2}$ and
$d *H = -g^2\sum_k F_k \wedge * F_{k+2}$. However, both 
have been shown to be implied by the supersymmetry equations, in \cite{gmpt3} and \cite{koerber-tsimpis,koerber-martucci-AdS}
respectively (for all supersymmetric vacua, and not only for the class  of AdS \sut\ structure vacua reviewed here). So we can forget about them and impose (\ref{eq:bianchi})
alone. Since ${\rm Re} \Omega \wedge J=0$, the only non--trivial 
case is $k=2$. We get
\begin{equation}
	\label{eq:dw2}
	d W_2^-= \frac23\Big(\tilde m^2-15 m^2\Big) {\rm Re} \Omega\ .
\end{equation}

Summing up, we have reviewed in this section the conditions
for an AdS$_4$ vacuum with internal \sut\ structure: they are 
given by equations (\ref{eq:susy}) and (\ref{eq:dw2}). 
It would be rather easy to find solutions to (\ref{eq:susy})
alone; the real problems come when trying to solve 
(\ref{eq:dw2}) as well. We will now review a family of \sut\
structures for which it is possible to compute $dW_2$, and then 
show in section \ref{sec:finding} that some of them support
string vacua.


\section{Geometry of twistor spaces} 
\label{sec:geo}

The twistor bundle on a manifold $M_k$ of dimension $k$ 
is the bundle of all almost complex structures compatible
with a metric on $M_k$. The fibre is hence given by 
${\rm SO}(k)/{\rm U}(k/2)$. For $k=4$, this is 
${\rm SO}(4)/{\rm U}(2)=\cc\pp^1=S^2$.
Hence the twistor bundle on a four--manifold is an $S^2$
fibration; its total space ${\rm Tw}(M_4) $ has dimension 6. 

We will now review some aspects of this fibration: 
its topology, complex structures and metrics. 

\subsection{Topology} 
\label{sub:topology}

For the topology, we will first focus on the case in which 
 $M_4=S^4$. It can be shown then that the total space of the
twistor fibration is actually $\cc\pp^3$: 
\begin{equation}
	\label{eq:cp3fibr}
	\xymatrix{ 
	S^2\ \ar@{^{(}->}[r] &\cc\pp^3\ar[d]&\hspace{-1cm}={\rm Tw}(S^4)\\
	&S^4&}\ .	
\end{equation}
One way to see this is to think of $S^4$ as of the quaternionic
projective line ${\Bbb H}\pp^1$. Then the projection map can be given 
as 
\begin{equation}\label{eq:proj}
	\cc\pp^3\ni (z_1,z_2,z_3,z_4)\,\buildrel p\over\mapsto\, (z_1 + j z_2,z_3+j z_4)
	\in {\Bbb H}\pp^1 \ . 
\end{equation}

$\cc\pp^3$ has Betti numbers $b_0=b_2=b_4=b_6=1$ and $b_1=b_3=b_5=0$.
In terms of the fibration (\ref{eq:cp3fibr}), the two--cycle is
just the fibre. One might get confused, however, in trying to 
identify the four--cycle. The twistor fibration cannot in this
case have a global section, because that would be a globally defined
almost complex structure on $S^4$, and it is known that none
exists. So the base cannot be literally used as a cycle. 

The answer can be found by looking at the map $p$ in (\ref{eq:proj}). 
Think of a hyperplane $\cc\pp^2\subset \cc\pp^3$ as the
union $\cc^2\cup \cc\pp^1$, where $\cc\pp^1$ is the line at infinity
of the projective plane $\cc\pp^2$. Then, the projection map 
$p$ is one--to--one on $\cc^2$, but projects $\cc\pp^1 $ to a point.
The result is a one--point compactification of $\cc^2$, which is
topologically $S^4$.

So far we have looked at ${\rm Tw}(S^4)=\cc\pp^3$. Although we will
devote less attention to it, there is another manifold to which 
the computations of section \ref{sec:finding} apply, namely 
${\rm Tw}(\cc\pp^2)$. In that case, the fibration is 
\begin{equation}
	\label{eq:flagfibr}
	\xymatrix{ 
	S^2\ \ar@{^{(}->}[r] &
	\frac{{\rm SU}(3)}{{\rm U}(1)\times {\rm U}(1)}\ar[d]&\hspace{-1cm} ={\rm Tw}(\cc\pp^2)\\
	&\cc\pp^2&}\ .	
\end{equation}
Another notation used for the total space so obtained is ${\Bbb F}(1,2;3)$; it is also often called ``flag manifold''. 
It is the space of complex planes and lines in $\cc^3$ such that
the line belongs to the plane. (The line is the ``pole'' and the plane is the ``flag''.) In equations: 
\begin{equation}\label{eq:f123}
	{\Bbb F}(1,2;3)=\Big\{(z^i,\tilde z^i)\in \cc\pp^2\times\cc\pp^2\ \ 
	{\rm such\ that}\ \
	\sum_{i=1}^3 z^i \tilde z^i=0\Big\}\ .
\end{equation}
One can fibre this space over either of the two $\cc\pp^2$ factors, 
by the map that forgets either the $z^i$ or the $\tilde z^i$. The
fibre is a $\cc\pp^1$. Finally, one can use for example the Gysin exact sequence to compute that the Betti numbers are $b_0=b_6$, $b_1=b_3=b_5=0$, $b_2=b_4=2$. Intuitively, the two two--cycles are
the $\cc\pp^1$ in each of the $\cc\pp^2$ in  (\ref{eq:f123}).


\subsection{Almost complex structures} 
\label{sub:acs}

Having clarified somewhat the topology of this fibration, we now
look at what almost complex structures can be defined
 on the total space
${\rm Tw}(M_4)$, going back to a general $M_4$. 
Let the twistor fibre have coordinates $\sigma^i$, $i=1,2,3$, 
so that $\sum (\sigma^i)^2=1$. 
 Since it is by definition the space of almost complex
structure compatible with a given metric, we can  
``tautologically'' write $I_4(\sigma^i)$, which means that there
is an almost complex structure $I_4$ on the base $M_4$ for any
choice of the coordinates $\sigma^i$ on the fibre. This is by 
definition a tensor on the total space of the fibration ${\rm Tw}(M_4)$. We cannot call it an almost complex structure on
${\rm Tw}(M_4)$, however, because it has rank four. To promote
it to rank six, we have to choose an action on vectors along the
fibre; since the fibre is $S^2$, we can take the usual Riemann complex
structure $I_2$ on it (explicitly, $I_2(\del_{\sigma^i})= 
\epsilon^{ijk}\sigma^j\del_{\sigma^k}$). So we can now combine 
the two in an almost complex structure on ${\rm Tw}(M_4)$. 
On a local basis of vectors, 
\begin{equation}\label{eq:twcs}
	\tilde I= 
	\left(\begin{array}{cc}\vspace{.2cm}
		I_2 & 0 \\ 
		0& I_4(\sigma)
		\end{array}\right)\ .
\end{equation}
Actually, we could have also combined them with a different
sign: 
\begin{equation}\label{eq:atwcs}
	I= 
	\left(\begin{array}{cc}\vspace{.2cm}
		-I_2 & 0 \\ 
		0& I_4(\sigma)
		\end{array}\right)\ .
\end{equation}
The difference between these two almost complex structures 
$\tilde I$ and $I$
on the total space was stressed in \cite{salamon,eells-salamon}. 
The first, $\tilde I$, is the most popular one because it is
integrable (namely, it is a complex structure, and not just 
an ``almost'' complex structure) whenever 
\cite{atiyah-hitchin-singer} the anti--self--dual
part of the Weyl tensor $W_-$ \footnote{This is
not to be confused with the form $W_2^-$ in (\ref{eq:susy}).} 
of $M_4$ is zero -- that is, when $M_4$ is self--dual. 
In contrast, $I$ is never integrable. But it
has a nice feature of its own: its first Chern class is actually zero. 

To highlight the difference, let us look at the particular case
(\ref{eq:cp3fibr}) once again. $\tilde I$ is the usual complex structure for $\cc\pp^3$; it has $c_1=4$, and so in particular
there is no globally defined  $(3,0)$--form for it (let alone
one in cohomology). This complex structure does not look very promising
for us, because there is no $\Omega$, but also because of another
fact. If one {\it does} have a $(3,0)$--form for an {\it almost} complex structure, the latter is integrable if and only if 
\begin{equation}\label{eq:int}
	(d \Omega)_{2,2}=0 \ .
\end{equation}
Looking back at (\ref{eq:susy}), we see that the almost complex structure we are looking for is only integrable if $W_2^-=0$ and
$\tilde m=0$. Looking at (\ref{eq:dw2}), we also conclude that $m=0$, and hence all fluxes are zero, the manifold is a \cy\ (see (\ref{eq:cy})), and the cosmological constant is zero (see (\ref{eq:cc})). So an integrable complex structure would take us
back to the usual \cy\ compactifications. 

So there are good reasons to focus on $I$ instead, which has 
$c_1=0$ (hence a globally defined $(3,0)$--form $\Omega$ exists)
and which is not integrable.

\subsection{\sut\ structure} 
\label{sub:su3}

To make progress, we need to complement the complex structure
and its associated $\Omega$ with a two--form $J$ that forms 
an \sut\ structure with it. This is always possible (because $\Omega$ alone
defines a ${\rm Sl}(3,\cc)$ structure, and ${\rm Sl}(3,\cc)$ 
is homotopically equivalent to $U(3)$). Explicitly, let us introduce a holomorphic vielbein $e^a$, $a=1,2,3$,
namely a basis of one--forms such that
\begin{equation}\label{eq:Iac}
	I^t e^a = i e^a \ . 
\end{equation}
(The transposition $^t$ is because $I$ is acting on
one--forms.)
More specifically, let us take $e^3$ along the fibre, and 
$e^{1,2}$ to be pullback of forms on the base. Hence we also 
have
\begin{equation}\label{eq:tIac}
	\tilde I^t e^{1,2}=i e^{1,2}\ ,\qquad \tilde I^t e^3= -i e^3\ .
\end{equation}

 In the case in 
which $M_4$ is self--dual (as defined above) and Einstein, 
\cite{xu} showed that
\begin{equation}\label{eq:xu}
	d 
	\left(\begin{array}{c}\vspace{.2cm}
		e^1\\ \vspace{.2cm} e^2 \\ e^3
	\end{array}\right)
	= \left(\begin{array}{cc}\vspace{.2cm}
		- \alpha&\\
		& {\rm Tr} (\alpha)
	\end{array}\right)\wedge
	\left(\begin{array}{c}\vspace{.2cm}
		e^1\\ \vspace{.2cm} e^2 \\ e^3
	\end{array}\right)
	+\frac1R\left(\begin{array}{c}\vspace{.2cm}
		\overline{e^2}\wedge \overline{e^3}\\\vspace{.2cm}
		\overline{e^3}\wedge \overline{e^1}\\
		\sigma\, \overline{e^1}\wedge \overline{e^2}
	\end{array}\right)\ .	
\end{equation}
Here, $\alpha$ is an antihermitian $2\times 2$ matrix of one--forms 
($\alpha_{ij}+\overline{\alpha_{ji}}=0 $) that
acts on $e^{1,2}$, $R$ is an overall length scale, 
and $\sigma$ parameterizes 
the curvature of $M_4$ relative
to the one of the fibre $S^2$, as we will see more explicitly later.

The reason (\ref{eq:xu}) is useful is that it allows us to check
explicitly the properties of $J$ and $\Omega$ that we need. 
Let us define the \sut\ structure and metric
\begin{equation}
	\label{eq:su,g}
	J=\frac i2 e^i \wedge\overline{e^i}\ ,\qquad
	\Omega= i e^1\wedge e^2 \wedge e^3 \ ,\qquad
	g_6= e^i \overline{e^i}\ .
\end{equation}
(The metric is actually determined by the \sut\ structure
$(J,\Omega)$, since \sut$\subset {\rm SO}(6)$). 

It is easy, then, to use (\ref{eq:xu}) to compute\vspace{.2cm}
\begin{equation}\vspace{.2cm}\label{eq:ws}
	\begin{picture}(50,20)(-25,0)
		\put(-20,-25){\line(1,0){340}}
		\put(-20,30){\line(1,0){340}}
		\put(-20,-25){\line(0,1){55}}
		\put(320,-25){\line(0,1){55}}
	\end{picture}\hspace{-1cm}
	\begin{array}{c}\vspace{.4cm}
		d J=-\frac1R (\sigma+2) {\rm Re} \Omega \ ,\qquad	
		d \Omega=i(W_2^-\wedge J + \frac2{3R}(\sigma+2)J^2) \ ,\\
		W_2^-=\frac2{3R} i(\sigma-1)(e^1\wedge \overline{e^1}+ e^2 \wedge
		\overline{e^2}-2 e^3\wedge \overline{e^3})\ ;	
	\end{array}
	\label{eq:master}
\end{equation}
notice that $W_2$ is $(1,1)$ and primitive with respect to $J$. One
can also compute \vspace{.1cm}
\begin{equation}\vspace{.2cm}\label{eq:dw2tw}
	\begin{picture}(50,20)(-30,0)
		\put(-20,-18){\line(1,0){150}}
		\put(-20,23){\line(1,0){150}}
		\put(-20,-18){\line(0,1){41}}
		\put(130,-18){\line(0,1){41}}
	\end{picture}\hspace{-1cm}
	dW_2^-= \frac8{3R^2} (\sigma-1)^2 {\rm Re} \Omega \ .
\end{equation}
These equations will become useful in the next section, to solve
(\ref{eq:susy}) and (\ref{eq:dw2}).\footnote{While this paper
was in preparation, 
the paper \cite{aldazabal-font} worked out 
many geometrical details about
existing solutions on $\cc\pp^3$. With a little more work along
those lines, one
can actually use their computations to the case with general $\sigma$, and
find (\ref{eq:ws}) and (\ref{eq:dw2tw}) in an alternative way.
Their parameters are then mapped to ours as 
$\sigma=2 \lambda^2$, $R=2 \lambda$.} 

As a cross--check of (\ref{eq:xu}), we can also define {\it locally}
a three--form $\tilde \Omega= i e^1\wedge e^2 \wedge \overline{e^3}$
for the complex structure (\ref{eq:twcs}) (compare (\ref{eq:tIac})), 
and a two--form $\tilde J= (i/2)(e^1\wedge \overline{e^1}+ e^2 \wedge
\overline{e^2}+ \overline{e^3}\wedge e^3)$. One gets
\begin{equation}
	d \tilde\Omega=2 \tilde\Omega\wedge{\rm Tr}\alpha
\end{equation}
 which implies that $(d \tilde \Omega)_{2,2}=0$, in agreement
with our earlier statement (see (\ref{eq:int}))
that $\tilde I$ is integrable when $M_4$ is self--dual.
When $\sigma=2$, one can also see that $d\tilde J=0$, which 
reproduces the fact that $\cc\pp^3$ admits a \ka\ metric.  
(We will see later again how the value $\sigma=2$ is special.)

In fact, if on top of the assumptions already made on $M_4$
to derive (\ref{eq:xu}) (namely, that $M_4$ be 
self--dual and Einstein) we also impose that it have positive scalar
curvature, we are left with only two nonsingular examples: 
$S^4$ and $\cc\pp^2$ (see for example \cite{besse}). Even if we do not need to 
restrict to $\sigma=2$ (we are not using the complex structure
$\tilde I$, after all, nor do we want ${\rm Tw}(M_4)$ to be
\ka), we will see in section \ref{sec:finding} that 
we still need $\sigma>0$, so that we will only be left 
with $\cc\pp^3$ and ${\Bbb F}(1,2;3)$.


\subsection{Metric} 
\label{sub:metric}
Both almost complex structures $I$ and $\tilde I$ are compatible
with the same metric defined in (\ref{eq:su,g}). We end this
section by reviewing some features of this metric.
We can take the relevant computations from \cite[Sec.~1]{gibbons-page-pope}\footnote{For $R=1$, 
$\lambda_{\rm GPP}=\sqrt{\sigma/2}$.}. Let us define
\begin{equation}
	g_4 \equiv e^1 \overline{e^1}+ e^2 	\overline{e^2}\ ,\qquad
	g_2 \equiv e^3 \overline{e^3}\ . 
\end{equation}
Then we have\footnote{One could also derive these formulas 
directly from the machinery of \sut\ structures, as for example
in \cite{bedulli-vezzoni}; the Ricci scalar is particularly easy
to cross--check in this way.}
\begin{equation}\label{eq:ricci}
	{\rm Ric}_4= \frac{\sigma(6-\sigma)}{R^2}g_4 \ ,\qquad	
	{\rm Ric}_2= \frac{\sigma^2+4}{R^2}g_2	\ .
\end{equation}
We see that the metric $g$ is Einstein if and only if
\begin{equation}
	\sigma=1 \qquad {\rm or }\qquad \sigma=2 \qquad {\rm (Einstein).}
\end{equation}
We will see that both these cases have already been used to
construct vacua (in \cite{nilsson-pope,watamura-2,sorokin-tkach-volkov-11-10} and \cite{behrndt-cvetic,behrndt-cvetic2},
respectively). 


\section{Finding vacua} 
\label{sec:finding}

We now have all the ingredients we need to solve
the supersymmetry equations (\ref{eq:susy}) and Bianchi
identities (\ref{eq:dw2}). If we do, we will have found 
a IIA supergravity solution. We will first do so, and then worry
about possible string theory corrections. 

\subsection{Supergravity} 
\label{sub:sugra}
We argued in section \ref{sub:su3} that a good candidate
for a flux vacuum is the twistor space
 ${\rm Tw}(M_4)$, when $M_4$ is self--dual and Einstein. 
Specifically, we proposed the almost complex structure $I$ 
given in (\ref{eq:atwcs}); and we derived in (\ref{eq:ws}) 
and (\ref{eq:dw2tw}) some relevant geometrical quantities. 
The \sut\ structure and the metric depend on a squashing
parameter $\sigma$, and on the overall scale $R$.

First of all, by comparing $dJ$ in (\ref{eq:ws}) with 
(\ref{eq:susy}), we get 
\begin{equation}\label{eq:tm}
	\tilde m = -\frac1{2R} (\sigma+2)\ .
\end{equation}
Next, comparing $d W_2^-$ in (\ref{eq:dw2tw}) with (\ref{eq:dw2}), 
we get, after some manipulation,
\begin{equation}\label{eq:m}
	m= 
	\frac1{2R}\sqrt{\left(\sigma-\frac25\right)(2-\sigma )}
	\equiv\frac1{2R}m_0(\sigma)
	\ ;
\end{equation}
in particular,
\begin{equation}\label{eq:interval}
	\begin{picture}(50,20)(-25,0)
		\put(-15,-20){\line(1,0){80}}
		\put(-15,25){\line(1,0){80}}
		\put(-15,-20){\line(0,1){45}}
		\put(65,-20){\line(0,1){45}}
	\end{picture}\hspace{-1cm}
	\frac25 \le \sigma \le 2\ . 
\end{equation}
\vspace{.2cm}

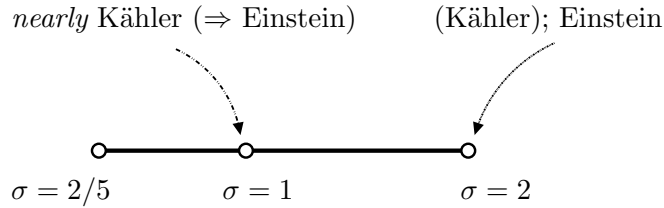
\begin{figure}[h]
\begin{picture}(200,80)(0,-20)
	\put(150,0){\includegraphics[width=15em]{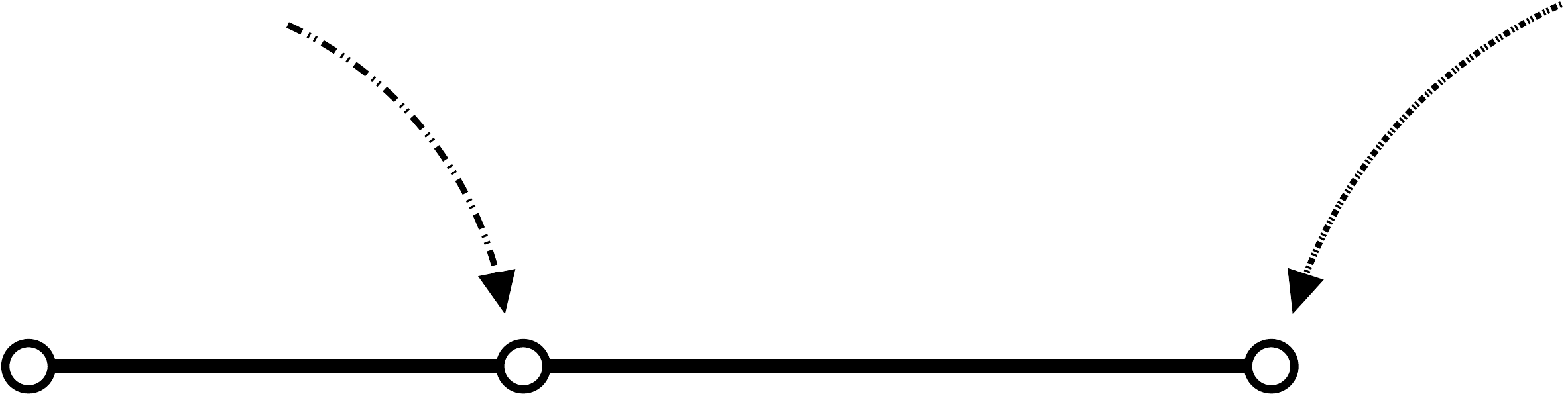}}
	\put(120,50){\small {\it nearly} \ka\ ($\Rightarrow$ Einstein)}
	\put(280,50){\small (\ka); Einstein}
	\put(120,-15){\small $\sigma=2/5$}
	\put(200,-15){\small $\sigma=1$}
	\put(290,-15){\small $\sigma=2$}
\end{picture}
\caption{\small This sketch shows the allowed interval for $\sigma$ in 
(\ref{eq:interval}), along with the three special cases already
used for string vacua before this paper. This is {\it not} a moduli
space, because of flux quantization, as discussed in section 
\ref{sub:quant}. In the two extrema
\cite{nilsson-pope,watamura-2,sorokin-tkach-volkov-11-10}, the Romans mass vanishes (see (\ref{eq:m}) and
(\ref{eq:susy})); the solution can hence be lifted to M--theory. 
The resulting seven--dimensional metric on $S^7$ is Einstein in 
both cases.
The metric at $\sigma=2$ admits a \ka\ structure,
but supersymmetry uses another almost complex structure. The case
$\sigma=1$ was used in \cite{behrndt-cvetic,behrndt-cvetic2}.}
\end{figure}

Since $\sigma$ then has to be positive, we have (as commented
at the end of \ref{sub:su3}) that the only two manifolds on which
we can apply the methods of this paper are $\cc\pp^3$ and ${\Bbb F}(1,2;3)$. On each of these, however, we will find infinitely many
vacua. 

At this point, as far as IIA supergravity is concerned, we are done. 
We have satisfied the equations for $dJ$ and $d \Omega$ in
 (\ref{eq:susy}), and the one for $d W_2^-$ in (\ref{eq:dw2}), by
taking the parameters $\tilde m$ and $m$ to be given by 
(\ref{eq:tm}) and (\ref{eq:m}). The fluxes are then given in
(\ref{eq:susy}). 
 
We also know from the general
theory (as commented in section \ref{sec:review}) that the equations
of motion will be automatically satisfied. It is also not difficult
to check them directly, by using the expressions for the fluxes in
(\ref{eq:susy}) and (\ref{eq:ricci}). 

Since we want, however, to find {\it string theory} vacua and
not just supergravity solutions, we have to now turn to flux
quantization effects, and to possible stringy corrections.


\subsection{Flux quantization} 
\label{sub:quant}

The fluxes in (\ref{eq:susy}) cannot be quantized. $H$ is actually
exact:
\begin{equation}\label{eq:H}
	H=d \left(\frac m{\tilde m} J \right)\ ,
\end{equation}
so its periods are zero;
as for the $F_k$, they are not closed, because of 
(\ref{eq:bianchi}), and hence their periods are
not well--defined. Thanks to (\ref{eq:H}), however, 
we can define ``Page charges''
\begin{equation}\label{eq:B}
	\tilde F_k\equiv e^{-B\wedge}F_k \ ,\qquad
	B=\frac m{\tilde m} J+ B_0\ ,
\end{equation}
where $B_0$ is closed. We will set $B_0$ to zero in what follows, 
since this choice will be enough for finding vacua. We can
then compute explicitly: 
\begin{equation}\label{eq:fgr}
	\begin{array}{c}\vspace{.3cm}
		\tilde F_2=\frac 4{g_s R}\frac{(\sigma-1)}{(\sigma+2)}
		\left[-\sigma j_4+ 2 j_2\right]\ ,\qquad
		\tilde F_4=-2\frac{m_0}{g_s R(\sigma+2)^2}
		\left[(\sigma^2-2 \sigma-2) j_4^2-6\sigma j_2\wedge j_4\right]\ ,\\
		\tilde F_6=\frac 8{15 g_s R}\frac{(1+2 \sigma)(-\sigma^2+12 \sigma +4)}{(\sigma+2)^3}\, J^3\ ,	
	\end{array}	
\end{equation}
where $m_0(\sigma)$ has been defined in (\ref{eq:m}) and
\begin{equation}
	j_4=\frac i2(e^1 \wedge \overline{e^1}+e^2 \wedge \overline{e^2})
	\ ,\qquad
	j_2=\frac i2\,e^3 \wedge \overline{e^3}\ .
\end{equation}

We can now impose flux quantization. To some extent, the proper
understanding of what this means is still work in progress
(see for example \cite{belov-moore}). For example, which fluxes
are quantized depends on our choice of electric basis of field--strengths; the choice should cancel in the partition function,
but it does matter when trying to decide whether a single 
given configuration is a solution or not. In the present
situation, the wisest course of action would seeem to just
impose that the internal fluxes $F_k$ be quantized according
to the formula ${\rm ch}(x)\sqrt{\hat A}$, where $x$ is an element of
the K--theory group \cite{moore-witten}. This formula gives rise to several subtleties, such as $F_6$ being actually half--integral
or integral depending on the value of $F_4$. Working this
out carefully seems to be beyond the scope of the present paper, since,
as we will see, it does not affect the existence of solutions. 
We will impose, schematically, 
\begin{equation}\label{eq:qf}
	\int_{C_a}\tilde F_k= n_k(2\pi l_s)^{k-1}
\end{equation}
on all the internal $F_k$. To fix 
ideas, we can keep in mind the ``naive'' reduction of the
half--quantization of \cite{witten-fivebrane} from M--theory. Lastly, notice that allowing a non--zero $B_0$ (as
in (\ref{eq:B})) rather than setting it to zero as we did, 
will allow us even more freedom in the quantization, since it will
alter the formula in \cite{moore-witten} to ${\rm ch}(x)e^{B_0}\sqrt{\hat A}$.

On $\cc\pp^3$ there are four equations to be imposed. 
In $\tilde F_2$, the relevant term is the second, that integrates
on the fibre. In $\tilde F_4$, it is the first term that we are interested in: as we remarked in \ref{sub:topology}, even if the
base is not a cycle, a $\cc\pp^2\subset \cc\pp^3$ projects to 
the base by collapsing the line at infinity. 

After imposing (\ref{eq:qf}), from the equations for $n_4$
and $n_0$ one can derive:
\begin{equation}\label{eq:gr}
	\begin{array}{c}\vspace{.3cm}
		g_s=n_0^{-3/4} n_4^{-1/4}
		\frac{m_0(\sigma)}{\sqrt{\sigma (\sigma+2)}}\left(
		\frac{125 \pi^2}{6 }		
		(1-\sigma)(2 \sigma +1 )\right)^{1/4}\\
		\frac{R}{2\pi l_s}\equiv
		r=n_0^{-1/4} n_4^{1/4}\sqrt{\sigma(\sigma+2)}\left(
		\frac{8\pi^2}{15}(1- \sigma)(2 \sigma +1 )
		\right)^{-1/4}\ . 
	\end{array}
\end{equation}
We chose to derive $g_s$ and $r$ from the equations for $n_0$ and $n_4$ because the functions of $\sigma$ that they contain 
are both positive and bounded within the allowed interval (\ref{eq:interval}); this will be useful shortly. 
In particular, we have $n_0>0$ and $n_4>0$. (We will assume 
from now on that $\sigma$ is not one of the special values
0, 1 or $2/5$.) 

We can then determine $\sigma$ by
\begin{equation}\label{eq:n2}
	\frac{n_2}{\sqrt{n_0 n_4}}= \sqrt{\frac{24}{5}}\frac{\sigma}{m_0(\sigma)}
	\sqrt{\frac{\sigma -1 }{2 \sigma +1 }}\ 
\end{equation}
and there is one $\sigma$ in the allowed interval (\ref{eq:interval}) for any integer $n_2$, negative or positive. 
So we have now fixed the three moduli $g_s$, $\sigma$ and $r$ in
terms of the (half)integers $n_{0,2,4}$. It would seem, however, that
we are going to run into trouble when we impose the quantization of
$F_6$. Fortunately, the relevant equation is
\begin{equation}\label{eq:n6}
	n_6=\left(\frac{n_2 n_4}{n_0}\right)\frac{\sigma^2 -12 \sigma -4}{8 (\sigma -1)^2}\ ;
\end{equation}
the fact that the function on the right hand side is a rational
function with rational coefficients is what saves us. Here is why. 
Let us first of all restrict our attention to $\sigma$ rational. 
Before (\ref{eq:n6}), one can choose any $n_{0,2,4}$ and 
determine $g_s, r, \sigma$.
Let us now give up a bit of that freedom, and choose $n_{0,4}$ so that they cancel the square root
that will appear in the denominator of the function on the right hand
side of (\ref{eq:n2}). So far we have three particular integers
$n_0^0, n_2^0, n_4^0$ of (\ref{eq:gr})
and (\ref{eq:n2}) with $\sigma$ rational and some $g_s$ and $r$. 
Let us now look at the right hand side of (\ref{eq:n6}). It will read at this point $\frac{n_2^0 n_4^0}{n^0_0}\frac{N_1}{N_2}$, for some integers $N_1, N_2$
 (since $\sigma$ is rational). It is now sufficient to take
$n_{0,2,4}=(n_0^0 N_2) n_{0,2,4}^0$ (so that the solution 
for $\sigma$ to 
(\ref{eq:n2}) does not change; $g_s$ and $r$ will change, but so be it), and $n_6=n_2 n_4 N_1$. 

In the discussion so far, we have set $B_0$ in (\ref{eq:B}) to zero. Had we 
allowed it to be non--zero, we would have had one more parameter to 
vary (since on $\Bbb{CP}^3$ there is one harmonic two--form), which 
would have resulted in a system of four equations
for four unknowns. To find solutions to this more general system 
one clearly does not have to work as hard as we had to for 
$B_0=0$.\footnote{I thank Amir Kashani--Poor for discussions on this point.}

Once one has found a particular solution $\{ n_k=\bar n_k\}$, one
can find infinitely
many others by rescaling. While we are at it, we can choose the
rescaling so as to make $r$ parametrically large and $g_s$ parametrically
small: 
\begin{equation}\label{eq:rescale}
	\{n_0=N^2 \bar n_0\ , \ n_2=N^3 \bar n_2\ ,\ n_4=N^4 \bar n_4\ , \ 
	n_6=N^5 \bar n_6\}
\end{equation}
under which $\sigma$ remains invariant, $g_s\sim N^{-5/2}$, 
$r\sim N^{1/2}$. (We should take $N$ odd so that it preserves
any half--integrality.)  
Other choices of rescalings are of course possible. 

With this rescaling we have made sure that both $l_s$ and $g_s$
corrections are under control, but one might worry 
about the
fact that we are introducing ever larger quanta of flux. 
One might think that this would make large any corrections to the action in which
the flux appears with high powers, for example.  
As remarked in \cite{dewolfe-giryavets-kachru-taylor}, such
corrections should be functions of $s_k\equiv (g_s F_k)^2$, where the
 square
is actually a contraction of the indices, which involves $k$ inverse 
metrics. Suppose for example that we are looking at the behavior
of $s_k$ under the rescaling (\ref{eq:rescale}), so that
we can forget about the dependence on $\sigma$. 
The flux density (as opposed to the integral) $F_k$ goes
like $1/(g_s R)$, so $s_k$ goes like $(1/r)^2 r^{-2k}=r^{-2(k+1)}$, taking into account the $k$ inverse metrics. This means that 
$s_k$ is small when $r $ is large, and in particular that
it gets smaller under (\ref{eq:rescale}). 

The discussion for ${\Bbb F}(1,2;3)$ is very similar. Some numerical factors are different (essentially because
of the different metric on the base). 
More importantly, there is an additional two--cycle (coming from a $\cc\pp^1$ in 
the base) and an additional four--cycle (coming from the restriction
of the fibration to that $\cc\pp^1$). This might sound worrying, 
because we are then imposing two more equations. But in fact, if we
call $\tilde n_2$ and $\tilde n_4$ the two new integers, 
we can see from (\ref{eq:fgr}), with some work, that 
$\tilde n_2/ n_2$
and $\tilde n_4/n_4$ are rational functions with rational
coefficients. One can then perform the rescaling (\ref{eq:rescale})
until $\tilde n_{2,4}$ can be taken to be integer.


\subsection{Comments and possible extensions} 
\label{sub:comments}

Now that we have convinced ourselves that the supergravity
vacua found in \ref{sub:sugra} survive the gauntlet of flux 
quantization and possible stringy corrections, we can ask
whether they are in fact interesting physically. 

The first feature that springs to mind is the fact that at
this point there are no {\it known} moduli left. There were only
two geometrical moduli in our metric, $R$ and $g$, and they
have been stabilized along with the dilaton in section
\ref{sub:quant}. Often, additional moduli can come from potentials, 
but in this case the RR potentials are odd forms and have no cycles
to be integrated on; $B_0$ can be integrated on the two--cycle, but
it is not a modulus, since for example it shifts $F_2\to F_2 -B_0 F_0$, 
which does not respect (\ref{eq:qf}). It would have
been suspicious anyway if there had been moduli coming from potentials, since these moduli are typically supersymmetry partners of geometrical moduli, and one cannot stabilize a field and not its partner without breaking supersymmetry. 

Unfortunately, without having performed the whole KK reduction, 
we cannot be sure yet that there are no other moduli that we have
not thought about. It is not even enough to know the spectrum of
the Laplacian, because the internal fluxes mix with it (for an example, see \cite[Table 5]{duff-nilsson-pope}). 

It would be interesting at this point to know more about the mass
matrix around these vacua. For the subset found at
$\sigma=1$ by \cite{behrndt-cvetic,behrndt-cvetic2}, the effective theory for a subset of fields is now known \cite{kashanipoor}, and the masses are positive (not just about the stability bound). 

This would be interesting in view of a possible uplifting of these
vacua.\footnote{Notice that the no--go arguments in \cite{hertzberg-kachru-taylor-tegmark} about de Sitter vacua only applies to \cy\ spaces.} If one wants to uplift an AdS vacuum which has some masses
over the stability bound but negative, the uplifting term in the
potential is unlikely to make them positive unless it has itself
a minimum at the vacuum. Hence having positive masses from the
beginning appears desirable. 

The uplifting would hopefully also cure one unpleasant feature
of the vacua in this paper, that we have not remarked so far. 
Namely, there
is no separation of scales between the four--dimensional cosmological constant and the Kaluza--Klein scale. Indeed, from (\ref{eq:cc}),
and using (\ref{eq:m}), (\ref{eq:tm}),
\begin{equation}
	\Lambda R^2=-3(m^2 + \tilde m^2)R^2= \frac{12}5 (2 \sigma+1)
\end{equation}
whereas one would have liked this number to be small (it is proportional to $(H/m_{\rm KK})^2$, where $H$ is the Hubble scale).
This is unlike the vacua  in \cite{dewolfe-giryavets-kachru-taylor}, where, in the notation of this paper, $R\sim n_4/n_0$, 
$g_s\sim n_4^{-3/4}n_0^{-5/4}$, and $\Lambda R^2\sim g_s^2 R^2
\sim n_4^{-1}n_0^{-3}$. The crucial difference appears to be the presence, in their case, of orientifold sources; we will have
some speculative comments about this at the end of this section. 
In any case, as already mentioned, the position taken in this paper is that this kind of question should be asked only after
the uplifting.

Another question we are not answering regards the gauge group
of the effective theory. For example, the metrics considered here
for $\cc\pp^3$ have isometries
 ${\rm Sp}(2)$, but the fluxes might break
some of them, and mix the survivors with the vector in \cite{kashanipoor} (that
comes from the RR potential $A_3$) in a semi--direct product, similarly to what happens in Scherk--Schwarz reductions.\footnote{In trying
to understand better the supersymmetry of the effective action, 
the superspace constructions of \cite{grassi-marescotti} might 
turn out to be useful.}

It appears possible to answer all these questions with a
reasonable amount of work. In any case, the point of this paper
is less in the features of the vacua than in the techniques 
utilized 
to obtain them, that hopefully might become of more general use. 

Concretely, here is a more speculative possibility. 
So far we have not introduced any RR source, because
they usually make the equations much more difficult to solve. There
is a brutal approximation that in many cases seems to reproduce vacua
that one has otherwise good control on: it consists in replacing
the source say for an O6--plane, that would look locally like
\begin{equation}\label{eq:delta}
	-\mu\delta(x^1)\delta(x^2)\delta(x^3)dx^1\wedge dx^2\wedge dx^3\ ,
\end{equation} 
 $x^i$ being the transverse coordinates, with a non--singular form.
\cite{acharya-benini-valandro} proposes taking
\begin{equation}\label{eq:Re}
	-\frac{\mu}{R^3}{\rm Re} \Omega 
\end{equation}
with the obvious motivation that it would then be comparable to
the existing terms in (\ref{eq:dw2}).\footnote{Notice that,
with a non--smeared source such as in (\ref{eq:delta}), one would
expect a non--trivial warping and dilaton, whereas we have seen 
in section \ref{sec:review} that this is not possible for 
vacua with \sut\ structure. This presumably means that one has
to consider vacua with \stt\ structure instead.}
 One possible way to think
about it is to expand (\ref{eq:delta}) in eigenforms of the Laplacian,
and keep the lowest mode. 
 
In any case, if one believes in this approximation, one can try to
combine it with the computations in this paper. After adding
(\ref{eq:Re}) to the right hand side of (\ref{eq:dw2}), one finds
that (\ref{eq:tm}) gets modified to
\begin{equation}
	m= 
	\sqrt{\frac1{4R^2}\left(\sigma-\frac25\right)(2-\sigma )+
	\frac{g \mu}{10 R^3}}
	\ .
\end{equation}
As expected, the introduction of the O6--plane makes the
equations more forgiving: it becomes possible a priori to have
{\it negative} $\sigma$, which would correspond to the twistor space
of a hyperbolic $M_4$ (such as quotients of hyperbolic four--space).
It then also becomes possible to make $\sigma$ close to $-1/2$, which  
would introduce a hierarchy of scales between the four--dimensional
cosmological constant and the KK scale, as discussed above. 
However, we will not investigate further this possibility here.


{\bf Acknowledgments}. This work was supported in part by DOE grant DE-FG02-91ER4064. It is a pleasure to thank C.~Beasley, 
D.~Cassani, F.~Denef, M.~Gra\~na, S.~Kachru, A.--K.~Kashani--Poor, G.~Moore, A.~Moroianu,
G.~Policastro, G.~Villadoro, X.~Yin, F.~Xu  for interesting discussions and correspondence. 


\begin{thebibliography}{10}

\bibitem{maldacena-nunez}
J.~M. Maldacena and C.~Nu{\~n}ez, ``Supergravity description of field theories
  on curved manifolds and a no--go theorem,'' {\em Int. J. Mod. Phys.} {\bf
  A16} (2001) 822--855,
\href{http://arXiv.org/abs/hep-th/0007018}{{\tt hep-th/0007018}}.

\bibitem{kklt}
S.~Kachru, R.~Kallosh, A.~Linde, and S.~P. Trivedi, ``{De Sitter} vacua in
  string theory,'' {\em Phys. Rev.} {\bf D68} (2003) 046005,
\href{http://arXiv.org/abs/hep-th/0301240}{{\tt hep-th/0301240}}.

\bibitem{dasgupta-rajesh-sethi}
K.~Dasgupta, G.~Rajesh, and S.~Sethi, ``M theory, orientifolds and
  {$G$}--flux,'' {\em JHEP} {\bf 08} (1999) 023,
\href{http://arXiv.org/abs/hep-th/9908088}{{\tt hep-th/9908088}}.

\bibitem{grana-polchinski}
M.~Gra{\~n}a and J.~Polchinski, ``Supersymmetric {three--form} flux
  perturbations on {AdS$_5$},'' {\em Phys. Rev.} {\bf D63} (2001) 026001,
\href{http://arXiv.org/abs/hep-th/0009211}{{\tt hep-th/0009211}}.

\bibitem{giddings-kachru-polchinski}
S.~B. Giddings, S.~Kachru, and J.~Polchinski, ``Hierarchies from fluxes in
  string compactifications,'' {\em Phys. Rev.} {\bf D66} (2002) 106006,
\href{http://arXiv.org/abs/hep-th/0105097}{{\tt hep-th/0105097}}.

\bibitem{gmpt2}
M.~Gra{\~n}a, R.~Minasian, M.~Petrini, and A.~Tomasiello, ``Generalized
  structures of {${\cal N}=1$} vacua,'' {\em JHEP} {\bf 11} (2005) 020,
\href{http://arXiv.org/abs/hep-th/0505212}{{\tt hep-th/0505212}}.

\bibitem{t-reform}
A.~Tomasiello, ``{Reformulating Supersymmetry with a Generalized {Dolbeault}
  Operator},'' {\em JHEP} {\bf 02} (2008) 010,
\href{http://arXiv.org/abs/arXiv:0704.2613 [hep-th]}{{\tt arXiv:0704.2613
  [hep-th]}}.

\bibitem{gmpt3}
M.~Gra{\~n}a, R.~Minasian, M.~Petrini, and A.~Tomasiello, ``A scan for new
  {${\cal N}=1$} vacua on twisted tori,'' {\em JHEP} {\bf 05} (2007) 031,
\href{http://arXiv.org/abs/hep-th/0609124}{{\tt hep-th/0609124}}.

\bibitem{dewolfe-giryavets-kachru-taylor}
O.~DeWolfe, A.~Giryavets, S.~Kachru, and W.~Taylor, ``Type {IIA} moduli
  stabilization,'' {\em JHEP} {\bf 07} (2005) 066,
\href{http://arXiv.org/abs/hep-th/0505160}{{\tt hep-th/0505160}}.

\bibitem{acharya-benini-valandro}
B.~S. Acharya, F.~Benini, and R.~Valandro, ``Fixing moduli in exact type {IIA}
  flux vacua,'' {\em JHEP} {\bf 02} (2007) 018,
\href{http://arXiv.org/abs/hep-th/0607223}{{\tt hep-th/0607223}}.

\bibitem{duff-nilsson-pope}
M.~J. Duff, B.~E.~W. Nilsson, and C.~N. Pope, ``{Kaluza-Klein} supergravity,''
  {\em Phys. Rept.} {\bf 130} (1986)
1--142.

\bibitem{nilsson-pope}
B.~E.~W. Nilsson and C.~N. Pope, ``Hopf fibration of eleven-dimensional
  supergravity,'' {\em Class. Quant. Grav.} {\bf 1} (1984)
499.

\bibitem{watamura-2}
S.~Watamura, ``Spontaneous compactification and {${\Bbb C}{\Bbb P}^N$}: {${\rm
  SU}(3) \times {\rm SU}(2) \times {\rm U}(1)$}, {$\sin^2(\theta_{\rm W})$},
  {$g_3/g_2$} and {${\rm SU}(3)$} triplet chiral fermions in four dimensions,''
  {\em Phys. Lett.} {\bf B136} (1984)
245.

\bibitem{sorokin-tkach-volkov-11-10}
D.~P. Sorokin, V.~I. Tkach, and D.~V. Volkov, ``On the relationship between
  compactified vacua of {$d = 11$} and {$d = 10$} supergravities,'' {\em Phys.
  Lett.} {\bf B161} (1985)
301--306.

\bibitem{duff-lu-pope}
M.~J. Duff, H.~Lu, and C.~N. Pope, ``Supersymmetry without supersymmetry,''
  {\em Phys. Lett.} {\bf B409} (1997) 136--144,
\href{http://arXiv.org/abs/hep-th/9704186}{{\tt hep-th/9704186}}.

\bibitem{behrndt-cvetic}
K.~Behrndt and M.~Cvetic, ``General {${\cal N} = 1$} supersymmetric fluxes in
  massive type {IIA} string theory,'' {\em Nucl. Phys.} {\bf B708} (2005)
  45--71,
\href{http://arXiv.org/abs/hep-th/0407263}{{\tt hep-th/0407263}}.

\bibitem{behrndt-cvetic2}
K.~Behrndt and M.~Cvetic, ``General {${\cal N} = 1$} supersymmetric flux vacua
  of (massive) type {IIA} string theory,'' {\em Phys. Rev. Lett.} {\bf 95}
  (2005) 021601,
\href{http://arXiv.org/abs/hep-th/0403049}{{\tt hep-th/0403049}}.

\bibitem{manousselis-prezas-zoupanos}
P.~Manousselis, N.~Prezas, and G.~Zoupanos, ``Supersymmetric compactifications
  of heterotic strings with fluxes and condensates,'' {\em Nucl. Phys.} {\bf
  B739} (2006) 85--105,
\href{http://arXiv.org/abs/hep-th/0511122}{{\tt hep-th/0511122}}.

\bibitem{frey-lippert}
A.~R. Frey and M.~Lippert, ``{AdS} strings with torsion: Non--complex heterotic
  compactifications,'' {\em Phys. Rev.} {\bf D72} (2005) 126001,
\href{http://arXiv.org/abs/hep-th/0507202}{{\tt hep-th/0507202}}.

\bibitem{kashanipoor}
A.-K. Kashani-Poor, ``Nearly {Kaehler} reduction,''
\href{http://arXiv.org/abs/arXiv:0709.4482 [hep-th]}{{\tt arXiv:0709.4482
  [hep-th]}}.

\bibitem{house-palti}
T.~House and E.~Palti, ``Effective action of (massive) {IIA} on manifolds with
  {${\rm SU}(3)$} structure,'' {\em Phys. Rev.} {\bf D72} (2005) 026004,
\href{http://arXiv.org/abs/hep-th/0505177}{{\tt hep-th/0505177}}.

\bibitem{aldazabal-font}
G.~Aldazabal and A.~Font, ``A second look at {${\cal N} = 1$} supersymmetric
  {AdS$_4$} vacua of type {IIA} supergravity,''
\href{http://arXiv.org/abs/arXiv:0712.1021 [hep-th]}{{\tt arXiv:0712.1021
  [hep-th]}}.

\bibitem{ziller}
W.~Ziller, ``Homogeneous {Einstein} metrics on spheres and projective spaces,''
  {\em Math. Ann.} {\bf D72} (1982) 351--358.

\bibitem{villadoro-zwirner}
G.~Villadoro and F.~Zwirner, ``{${\cal N}=1$} effective potential from dual
  {type--IIA} {D6/O6} orientifolds with general fluxes,'' {\em JHEP} {\bf 06}
  (2005) 047,
\href{http://arXiv.org/abs/hep-th/0503169}{{\tt hep-th/0503169}}.

\bibitem{camara-font-ibanez}
P.~G. Camara, A.~Font, and L.~E. Ibanez, ``Fluxes, moduli fixing and
  {MSSM--like} vacua in a simple {IIA} orientifold,'' {\em JHEP} {\bf 09}
  (2005) 013,
\href{http://arXiv.org/abs/hep-th/0506066}{{\tt hep-th/0506066}}.

\bibitem{derendinger-kounnas-petropoulos-zwirner}
J.-P. Derendinger, C.~Kounnas, P.~M. Petropoulos, and F.~Zwirner,
  ``Superpotentials in {IIA} compactifications with general fluxes,'' {\em
  Nucl. Phys.} {\bf B715} (2005) 211--233,
\href{http://arXiv.org/abs/hep-th/0411276}{{\tt hep-th/0411276}}.

\bibitem{grana-louis-waldram}
M.~Gra{\~n}a, J.~Louis, and D.~Waldram, ``{Hitchin} functionals in {${\cal N} =
  2$} supergravity,'' {\em JHEP} {\bf 01} (2006) 008,
\href{http://arXiv.org/abs/hep-th/0505264}{{\tt hep-th/0505264}}.

\bibitem{kashanipoor-minasian}
A.-K. Kashani-Poor and R.~Minasian, ``Towards reduction of type {II} theories
  on {${\rm SU}(3)$} structure manifolds,'' {\em JHEP} {\bf 03} (2007) 109,
\href{http://arXiv.org/abs/hep-th/0611106}{{\tt hep-th/0611106}}.

\bibitem{gurrieri-louis-micu-waldram}
S.~Gurrieri, J.~Louis, A.~Micu, and D.~Waldram, ``Mirror symmetry in
  generalized {Calabi--Yau} compactifications,'' {\em Nucl. Phys.} {\bf B654}
  (2003) 61--113,
\href{http://arXiv.org/abs/hep-th/0211102}{{\tt hep-th/0211102}}.

\bibitem{koerber-martucci-ten-four}
P.~Koerber and L.~Martucci, ``From ten to four and back again: how to
  generalize the geometry,'' {\em JHEP} {\bf 08} (2007) 059,
\href{http://arXiv.org/abs/arXiv:0707.1038 [hep-th]}{{\tt arXiv:0707.1038
  [hep-th]}}.

\bibitem{cassani-bilal}
D.~Cassani and A.~Bilal, ``Effective actions and {${\cal N}=1$} vacuum
  conditions from {${\rm SU}(3)\times {\rm SU}(3)$} compactifications,'' {\em
  JHEP} {\bf 09} (2007) 076,
\href{http://arXiv.org/abs/arXiv:0707.3125 [hep-th]}{{\tt arXiv:0707.3125
  [hep-th]}}.

\bibitem{schwarz-chernsimons}
J.~H. Schwarz, ``Superconformal {Chern--Simons} theories,'' {\em JHEP} {\bf 11}
  (2004) 078,
\href{http://arXiv.org/abs/hep-th/0411077}{{\tt hep-th/0411077}}.

\bibitem{gaiotto-yin}
D.~Gaiotto and X.~Yin, ``Notes on superconformal {Chern--Simons}--matter
  theories,'' {\em JHEP} {\bf 08} (2007) 056,
\href{http://arXiv.org/abs/arXiv:0704.3740 [hep-th]}{{\tt arXiv:0704.3740
  [hep-th]}}.

\bibitem{lust-tsimpis}
D.~{L\"ust} and D.~Tsimpis, ``Supersymmetric {AdS$_4$} compactifications of
  {IIA} supergravity,'' {\em JHEP} {\bf 02} (2005) 027,
\href{http://arXiv.org/abs/hep-th/0412250}{{\tt hep-th/0412250}}.

\bibitem{gray-hervella}
A.~Gray and L.~Hervella, ``The sixteen classes of almost hermitian manifolds
  and their linear invariant,'' {\em Ann. di Mat. Pura ed Appl.(IV)} {\bf 123}
  (1980) 35.

\bibitem{chiossi-salamon}
S.~Chiossi and S.~Salamon, ``The intrinsic torsion of {${\rm SU}(3)$} and
  {$G_2$} structures,''
\href{http://arXiv.org/abs/math/0202282}{{\tt math/0202282}}.

\bibitem{koerber-tsimpis}
P.~Koerber and D.~Tsimpis, ``Supersymmetric sources, integrability and
  generalized- structure compactifications,''
\href{http://arXiv.org/abs/arXiv:0706.1244 [hep-th]}{{\tt arXiv:0706.1244
  [hep-th]}}.

\bibitem{koerber-martucci-AdS}
P.~Koerber and L.~Martucci, ``{D--branes} on {AdS} flux compactifications,''
\href{http://arXiv.org/abs/arXiv:0710.5530 [hep-th]}{{\tt arXiv:0710.5530
  [hep-th]}}.

\bibitem{salamon}
S.~Salamon, ``Harmonic and holomorphic maps,'' in {\em Geometry seminar ``Luigi
  Bianchi'' II---1984}, vol.~1164 of {\em Lecture Notes in Math.},
  pp.~161--224.
\newblock Springer, Berlin, 1985.

\bibitem{eells-salamon}
J.~Eells and S.~Salamon, ``Twistorial construction of harmonic maps of surfaces
  into four--manifolds,'' {\em Ann. Scuola Norm. Sup. Pisa Cl. Sci.} {\bf 4}
  (1985), no.~4, 12.

\bibitem{atiyah-hitchin-singer}
M.~Atiyah, N.~Hitchin, and I.~Singer, ``Self-duality in four-dimensional
  {Riemannian} geometry,'' {\em Proc. Roy. Soc. London Ser. A} {\bf 362}
  (1978), no.~1711, 425--461.

\bibitem{xu}
F.~Xu, ``{${\rm SU}(3)$}--structures and special lagrangian geometries,''
\href{http://arXiv.org/abs/math/0610532}{{\tt math/0610532}}.

\bibitem{besse}
A.~L. Besse, {\em Einstein Manifolds}.
\newblock Ergebnisse Der Mathematik Und Ihrer Grenzgebiete, 3. Folge.
  Springer-Verlag, 1987.

\bibitem{gibbons-page-pope}
G.~W. Gibbons, D.~N. Page, and C.~N. Pope, ``Einstein metrics on {$S^3$},
  {${\Bbb R}^3$} and {${\Bbb R}^4$} bundles,'' {\em Commun. Math. Phys.} {\bf
  127} (1990)
529.

\bibitem{bedulli-vezzoni}
L.~Bedulli and L.~Vezzoni, ``The {Ricci} tensor of {${\rm
  SU}(3)$}--manifolds,''
\href{http://arXiv.org/abs/math/0606786}{{\tt math/0606786}}.

\bibitem{belov-moore}
D.~Belov and G.~W. Moore, ``Holographic action for the self-dual field,''
\href{http://arXiv.org/abs/hep-th/0605038}{{\tt hep-th/0605038}}.

\bibitem{moore-witten}
G.~W. Moore and E.~Witten, ``Self-duality, {Ramond-Ramond} fields, and
  {K--theory},'' {\em JHEP} {\bf 05} (2000) 032,
\href{http://arXiv.org/abs/hep-th/9912279}{{\tt hep-th/9912279}}.

\bibitem{witten-fivebrane}
E.~Witten, ``Five-brane effective action in {M--theory},'' {\em J. Geom. Phys.}
  {\bf 22} (1997) 103--133,
\href{http://arXiv.org/abs/hep-th/9610234}{{\tt hep-th/9610234}}.

\bibitem{hertzberg-kachru-taylor-tegmark}
M.~P. Hertzberg, S.~Kachru, W.~Taylor, and M.~Tegmark, ``Inflationary
  constraints on type {IIA} string theory,''
\href{http://arXiv.org/abs/arXiv:0711.2512 [hep-th]}{{\tt arXiv:0711.2512
  [hep-th]}}.

\bibitem{grassi-marescotti}
P.~A. Grassi and M.~Marescotti, ``Flux vacua and supermanifolds,'' {\em JHEP}
  {\bf 01} (2007) 068,
\href{http://arXiv.org/abs/hep-th/0607243}{{\tt hep-th/0607243}}.

\end{thebibliography}

\providecommand{\href}[2]{#2}

\end{document}